\documentclass[nonacm, acmsmall]{acmart}

\usepackage[most]{tcolorbox}
\tcbuselibrary{listings}
\usepackage{booktabs}
\usepackage{graphicx}
\usepackage{colortbl} 
\usepackage[normalem]{ulem}
\useunder{\uline}{\ul}{}

\usepackage[utf8]{inputenc}
\usepackage{csquotes}
\renewcommand{\mkbegdispquote}[2]{\itshape}
\usepackage{cleveref}
\crefformat{section}{\S#2#1#3} %
\crefformat{subsection}{\S#2#1#3}
\crefformat{subsubsection}{\S#2#1#3}

\renewcommand\footnotetextcopyrightpermission[1]{} %
\AtBeginDocument{%
  \providecommand\BibTeX{{%
    \normalfont B\kern-0.5em{\scshape i\kern-0.25em b}\kern-0.8em\TeX}}}

\setcopyright{acmlicensed}
\copyrightyear{2018}
\acmYear{2018}
\acmDOI{XXXXXXX.XXXXXXX}

\acmConference[Conference acronym 'XX]{Make sure to enter the correct
  conference title from your rights confirmation emai}{June 03--05,
  2018}{Woodstock, NY}
\acmISBN{978-1-4503-XXXX-X/18/06}

\usepackage{booktabs}
\usepackage[multiple]{footmisc}
\usepackage{threeparttable}
\usepackage{multirow}
\usepackage{amsmath}
\usepackage{mathtools}
\usepackage{graphicx}
\usepackage{subcaption}
\usepackage{balance}
\usepackage{color}
\usepackage{wrapfig}
\usepackage{arydshln}
\usepackage{float}
\usepackage{hyphenat}

\usepackage[commandnameprefix=always, draft]{changes}

\begin{document}

\newcommand{\Sys}{\textsc{FairFare}}

\title[\Sys{}]{\Sys{}: A Tool for Crowdsourcing Rideshare Data to Empower Labor Organizers}

\author{Dana Calacci}
\authornote{Equal Contribution}
\email{dcalacci@princeton.edu}
\affiliation{%
  \institution{Penn State University}
  \country{USA}
  }
\author{Varun Nagaraj Rao}
\authornotemark[1]
\affiliation{%
  \institution{Center for Information Technology Policy, Princeton University}
  \country{USA}
  }
\email{varunrao@princeton.edu}
\author{Samantha Dalal}
\authornotemark[1]
\affiliation{%
  \institution{University of Colorado Boulder}
  \country{USA}
  }
\email{samantha.dalal@colorado.edu}
\author{Catherine Di}
\email{cd3993@princeton.edu}
\affiliation{%
  \institution{Princeton University}
  \country{USA}
  }
\author{Kok-Wei Pua}
\email{kp7662@princeton.edu}
\affiliation{%
  \institution{Princeton University}
  \country{USA}
  }
\author{Andrew Schwartz}
\email{andrew@cornflowerlabs.com}
\affiliation{%
  \institution{Cornflower Labs}
  \country{USA}
  }
  \author{Danny Spitzberg}
\email{daspitzberg@berkeley.edu}
\affiliation{%
  \institution{UC Berkeley}
  \country{USA}
  }
\author{Andrés Monroy-Hernández}
\email{andresmh@princeton.edu}
\affiliation{%
  \institution{Center for Information Technology Policy, Princeton University}
  \country{USA}
  }

\renewcommand{\shortauthors}{Calacci, Nagaraj Rao and Dalal, et al.}

\begin{abstract}

\begin{figure}[htb]
    \centering
    \includegraphics[width=\linewidth]{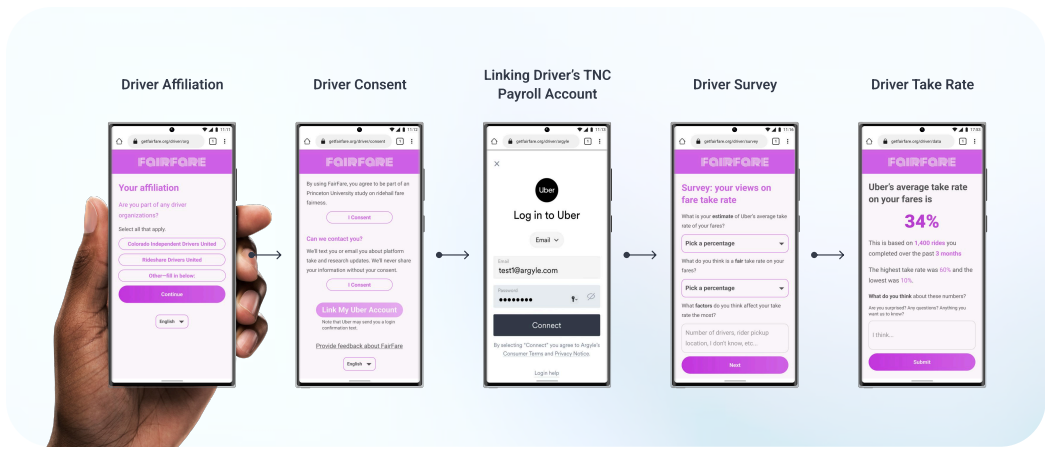}
    \caption{Screenshots of how drivers interact with the \Sys{} Web Application. Refer to \S \ref{sec:driver-interaction} for more details.}
    \label{fig:driver-interaction}
\end{figure}

Rideshare workers experience unpredictable working conditions due to gig work platforms’ reliance on opaque AI and algorithmic systems. In response to these challenges, we found that labor organizers want data to help them advocate for legislation to increase the transparency and accountability of these platforms. 
To address this need, we collaborated with a Colorado-based rideshare union to develop FairFare, a tool that crowdsources and analyzes workers' data to estimate the ``take rate'' --- the percentage of the rider price retained by the rideshare platform.
We deployed FairFare with our partner organization that collaborated with us in collecting data on 76,000+ trips from 45 drivers over 18 months. During evaluation interviews, organizers reported that FairFare helped influence the bill language and passage of Colorado Senate Bill 24-75, calling for greater transparency and data disclosure of platform operations, and create a ``national narrative''. Finally, we reflect on complexities of translating quantitative data into policy outcomes, nature of community based audits, and design implications for future transparency tools.

\end{abstract}

\keywords{}
\maketitle

\section{Introduction}

\label{sec:introduction}

Rideshare work can be financially precarious, physically dangerous, and psychologically harmful for workers~\cite{shapiroAutonomyControlStrategies2018a,rosenblatAlgorithmicLaborInformation2016b,rao2024transparency,dalal2023}. In recent years, labor organizers representing rideshare and delivery workers have advocated for regulations to improve working conditions in the rideshare industry that set wage floors and job loss protections~\cite{schwartzDeactivationRepresentationRole2023}. 
To call for these improvements, organizers need to understand workers' existing conditions~\cite{hookerMovingAlgorithmicBias2021}, a significant data access and social computing challenge in the rideshare industry. 

Labor organizers representing rideshare workers typically rely on a collage of qualitative anecdotes and screenshots to provide data about existing working conditions~\cite{dubalAlgorithmicWageDiscrimination2023}. While these qualitative data provide rich, ``thick descriptions'' ~\cite{geertz2008thick} of workers' experience, they are often dismissed by platforms as non-representative, cherry-picked examples. Rideshare platforms, on the other hand, have exclusive access to large-scale, comprehensive quantitative datasets of driver, trip, and pay data that they can draw upon to create authoritative narratives about working conditions in their industry~\cite{van2020platform}. Labor organizers need comprehensive access to large-scale quantitative data describing working conditions to conduct rigorous, independent investigations and contest platform-driven narratives.

There are tools and legal frameworks that empower \textbf{individual} rideshare workers to independently access quantitative work data (e.g., Gridwise and Data Subject Access Requests). However, these tools and frameworks do not provide an intuitive way to \textit{aggregate} individual worker data into a dataset that provides collective insight into overarching working conditions. Algorithmic auditing scholarship provides methods, like crowdsourcing data, to independently investigate black-boxed systems~\cite{sandvig2014auditing}. We build on efforts to develop tools for labor~\cite{fox2020worker} to design \Sys{}\footnote{\url{https://getfairfare.org/}}, a system that supports labor organizers in accessing aggregate, detailed, quantitative payment data for rideshare workers. \Sys{} is an initial step toward providing labor organizers with access to quantitative data describing pay, one metric of assessing working conditions in the rideshare industry.
\Sys{} is a digital tool that makes aggregating workers' historical work data easier for organizers by reducing the labor and risk involved in the collection. We designed and deployed \Sys{} with a labor organizing partner rather than individual gig workers to collect data that would specifically be useful for organizers' legislative advocacy efforts.

We focused on a specific issue that was informed by early initial engagement and relationship-building with our partner organization, a labor union based in Colorado, interested in influencing a local bill \cite{sb2475}. They identified \textit{unfairness} in the rideshare industry as a particular area where quantitative data could help their advocacy. We worked with our partner to operationalize this, and designed \Sys{} to measure \textit{take rates}: the proportion of a rider's fare that platforms take from each transaction before paying drivers.  \Sys{} is the result of nearly two years of collaboration with our partner organization, involving relationship-building, co-design, deployment, and ongoing evaluation. We continue to maintain \Sys{} today. In this paper, we describe how we designed, deployed, and evaluated \Sys{} as a tool to measure the impact of payment algorithms in rideshare work. \Sys{} was deployed with our partner organization in the summer of 2023. During the deployment, we conducted educational workshops for the partner organization's membership base to introduce the project, gather feedback, and address concerns about the design of \Sys{}. Our contributions include:  

\begin{itemize}
\item  Our collaborative research methodology for investigating rideshare platform take rates that was developed through iterative design with our partner organization and supports data-driven labor advocacy.
\item The design of \Sys{}, a tool for labor organizations to crowdsource historical rideshare trip and payment data from workers.
\item A field deployment of \Sys{} with our partner organization that demonstrates how it fits into their legislative advocacy process.
\item Post-deployment evaluation of the effectiveness and limitations of \Sys{} in policy advocacy based on feedback from organizers.
\end{itemize}

While \Sys{} tackles the crucial issue of independent quantitative data access that labor organizers face in the rideshare industry, it is just one piece of the larger puzzle in improving working conditions. Access to data is merely the first step toward supporting independent investigations that establish baseline measurements, which can then be used to advocate for necessary improvements. Our case study is specific to the gig economy, but we believe that the principles of participatory tool design and community collaboration may offer a template for broader applications. We also view these contributions as advancing policy-driven design through human-system interaction expertise addressing calls from recent work \cite{yang2024future}.

\section{Background and Related Work}

\label{sec:background-related-work}

\subsection{Gig Work, TNCs, and Data}

The rideshare economy, a significant sector of the gig economy, has become integral to U.S. transportation, with 36\% of adults using apps run by Transportation Network Companies (TNCs) such as Uber and Lyft to request rides, and 4.3\% of the workforce doing app-based work~\cite{axios2024app}. Despite the prevalence of the rideshare economy, detailed workforce data remain scarce~\cite{abrahamMeasuringGigEconomy2018,zipperer2022national,donovan2016does}. This data deficit hinders regulatory oversight~\cite{datadeficit}. 

Rideshare platforms leverage AI and algorithmically determined dynamic pricing structures that vary the price of rides and, consequently, payment to drivers in an unpredictable and opaque manner. Research demonstrates that dynamic pricing leads to financial and psychological harm for workers because of its unpredictable nature~\cite{shapiroAutonomyControlStrategies2018a,rosenblatAlgorithmicLaborInformation2016b}. Concerns about the discriminating effects of dynamic pricing have been raised~\cite{pandey2021disparate,chen2015peeking,yan2020dynamic}, yet independent, systematic investigation of discriminatory effects remains challenging. 

\subsubsection{Rideshare Transparency Policy Initiatives}
Transparency initiatives in the rideshare industry are advancing through legislative, regulatory, and grassroots efforts. Recent legislation, such as Washington State's HB 2076~\cite{wa2022transparency}, mandates disclosure of ride information, and support for deactivated drivers. Some cities, like New York and Chicago, now require data disclosures, enabling public oversight and revealing issues like price discrimination~\cite{pandey2021disparate}. However, these datasets remain geographically limited and lack details on gamified features and algorithmic operations. 
While researchers advocate for comprehensive public transparency reports \cite{rao2024ftt, rao2024memo, rao2024transparency}, drawing from social media transparency practices \cite{nagaraj2023discrimination, fb2024transparency} and generative AI model developers' calls for openness \cite{bommasani2024foundation}, such voluntary disclosures ultimately depend on companies' willingness to participate and their discretion in determining the scope and depth of transparency. 

Current legal efforts to support transparency initiatives are hampered by their reliance on the platform's willingness to comply in good faith with data disclosure requests and weak enforcement mechanisms. Further, lacking data, regulatory oversight is hard to instantiate. But without regulatory oversight right now, platforms don not have pressure to share that data. In the European Union, legal frameworks that enforce labor platform transparency have been implemented through Data Subject Access Requests (DSARs). Individual drivers can submit a DSAR to a rideshare platform, and theoretically, the platform would then share a copy of all data collected on that driver \cite{ausloos2018shattering, mahieu2020recognising}. The Worker Info Exchange (WIEX) in London attempted to leverage the legal framework provided by DSARs to collect and aggregate data on behalf of platform workers. However, this effort was hampered by the platforms' non-compliance with WIEX requests and a lack of legal recourse for data subjects facing platform non-compliance. While DSARs offer support for \textbf{individual} workers' access to data, the real value of worker data lies in the aggregate, not individual data points.

Complementing these top-down approaches, worker data collectives are emerging as powerful tools for empowerment\footnote{Worker data collectives are organizational structures that enable workers to pool and aggregate their work-related data to advance worker advocacy, enhance bargaining power, and improve accountability in the workplace \cite{hsieh2024worker}}. These collectives, which have been studied in HCI literature~\cite{imaginaries,uuapp,codesign} \cite{ zhang2024data}, encompass online and offline social institutions~\cite{atomized,la_strikes}, third-party tools for data sharing and analysis~\cite{uuapp,codesign,probes,self_track,sousveillance}, and platform evaluation mechanisms like Fairwork~\cite{fairwork}. Serving as communities of resistance~\cite{critique}, they enable collective data resistance strategies~\cite{leverage,refusal,strikes}.

Despite these transparency initiatives and the proliferation of data analytics tools intended to support individual drivers in gaining data-driven insights into their working patterns (e.g., Gridwise), no tools exist for the sole purpose of providing access to aggregate data about working conditions in app-based work for independent research. Moreover, no tools exist for \textbf{the sole purpose of supporting labor organizers} in the rideshare economy in collecting and analyzing aggregate datasets describing working and payment conditions. We attempt to bridge this gap by building a tool that enables labor organizers to collect and aggregate large amounts of historical data about rideshare working conditions and provide an evidentiary base for compelling advocacy.

\subsection{Tools and Strategies for Accessing Platform Data}

Platform workers have a long history of using digital tools to combat opaque algorithmic working conditions. Early efforts like Turkopticon~\cite{irani2013turkopticon} and Dynamo~\cite{salehiWeAreDynamo2015} enabled Amazon Mechanical Turk workers to collectively rate job posters, subverting the platform's one-sided rating system. Later tools, such as the Shipt Calculator~\cite{calacciBargainingBlackBoxDesigning2022}, attempted to reduce the human labor required for collective resistance by automating data collection and analysis. Individual resistance efforts have also emerged, with apps like Stopclub\footnote{\url{https://site.stopclub.com.br/en-us/en}}, Ubercheats\footnote{\url{https://radicaldata.org/projects/ubercheats/}}, Maxymo\footnote{\url{https://maxymoapp.com/}}, and Mystro\footnote{\url{https://www.mystrodriver.com/}} helping rideshare and delivery drivers calculate rates, identify payment discrepancies, and filter trips. These applications aim to address power asymmetries between individual workers and platforms by leveraging data.

Despite these advancements, there remains a gap in systematic data aggregation for investigating broader working conditions in app-based work. Historically, labor unions filled this role by maintaining data collection and analysis infrastructures to investigate management technologies' impacts and drive policy change~\cite{khovanskaya2020}. 
For instance, Rideshare Drivers United's worker-led study \cite{mccullough2022prop} analyzed data from 55 California drivers and 12,000 rides, revealing actual earnings of just \$6.20 per hour after expenses ---far below Prop 22's promises--- which helped catalyze driver protection laws in other states.
However, the decline of labor power\footnote{The term decline of labor power is often associated with a decline in \textbf{union} power in the US. However, here we mean to say that the overall decline in labor power more broadly refers to workers' limited ability to negotiate with employers due to a variety of factors which include, but are not limited to, the proliferation of fissured employment and subcontracting and computerization and automation. } in the US has diminished this capacity, reducing workers' ability to shape the future of algorithmically managed work \cite{kristal2013capitalist,greenbaum1996}. \Sys{} aims to address this data infrastructure gap by providing labor organizers with a tool to aggregate and analyze working conditions data in the rideshare sector, potentially revitalizing the labor movement's ability to engage in data-driven policy advocacy.

\subsection{Opportunities for Algorithmic Auditing}
Rideshare platforms rely on algorithmic decisions to determine key outcomes for drivers, including ride assignments and compensation. Algorithmic auditing, one approach for evaluating such opaque algorithmic systems, could provide workers and organizers with increased platform transparency and accountability ~\cite{sandvig2014auditing}. Current audit efforts focus on tool development for harm discovery, standards identification, performance evaluation, and results communication~\cite{ojewaleAIAccountabilityInfrastructure2024}. Algorithmic audits have been successfully used to identify discriminatory impacts of facial recognition systems~\cite{buolamwini2018gender} and reveal biases in job candidate screening algorithms~\cite{wilson2021building, nagaraj2023discrimination, ali2019discrimination}, resulting in concrete platform changes. However, studies examining the auditing landscape conclude that audits alone do not achieve accountability. Instead, they help investigate algorithms, which is just one step in the process \cite{ojewaleAIAccountabilityInfrastructure2024, ParticipationScaleTensions}. Furthermore, black-box audits' effectiveness depends heavily on companies' voluntary cooperation and good faith. Auditors often must rely on companies to provide historical data and deployment access to conduct thorough evaluations \cite{casper2024black}.

\subsubsection{Crowdsourced Data in Audits}

Crowdsourcing data, where many individuals voluntarily contribute data to a project, has become a crucial strategy for independent researchers and auditors. Researchers typically access platform data through two main approaches: (1) scraping, APIs, and public libraries, or (2) privileged access granted by platforms. However, platforms increasingly restrict access by increasing costs, banning researcher accounts~\cite{edelson2021we, bruns2021after}, and gatekeeping access to their data~\cite{breuer2020practical}. In response, crowdsourced data collection is a crucial strategy for independent data access and potential accountability. This approach works around platforms' API usage restrictions and can enable individual community members to more actively participate in the audit process\cite{d2024counting,currie2016police}.

Crowdsourced data collection methods range from voluntary data donation platforms like Mozilla Rally\footnote{\url{https://blog.mozilla.org/data/2021/05/05/announcing-mozilla-rally/}} to compensated participation~\cite{lukito2023enabling}. Data can be sourced through platform exports~\cite{razi2022instagram}, browser extensions~\cite{wojcieszak2022avenues}, or even screenshots~\cite{calacciBargainingBlackBoxDesigning2022}. 
Our work contributes to this space by %
exploring data donation that enables labor organizers to crowdsource historical rideshare and compensation data from drivers. This approach significantly reduces the human labor typically required for collecting and analyzing screenshots, addressing a key challenge in existing crowdsourced data efforts in the rideshare sector~\cite{khovanskaya2020}. 

\subsubsection{The Participatory Turn in Auditing}
A participatory approach in algorithmic auditing aims to address this accountability gap by involving impacted stakeholders in system development and monitoring~\cite{birhanePowerPeopleOpportunities2022,delgadoParticipatoryTurnAI2023}. Participatory audit approaches, primarily focused on harms discovery~\cite{lam2022end,hannakMeasuringPriceDiscrimination2014,attenberg2015,suh2019,ochigame2021}, improve end-user experience, but often limit participant agency in audit design and execution. This constraint, coupled with the risk of creating an epistemic burden on community groups when researchers ask them to produce data or engage in the production and upkeep of data for a research project~\cite{lam2023sociotechnical,pierreGettingOurselvesTogether2021}, highlights significant gaps in current participatory audit practices as well. Moreover, the existing audit landscape largely focuses on investigating algorithms but often falls short of translating findings into impact for affected communities. We collaborate with labor organizations to co-design an audit and translate technical audit results into a format that helps organizers achieve impact as they defined it--- crafting transparency legislation.

\section{Co-designing a Topic and Method of Inquiry}
Participatory approaches to algorithmic audits are critiqued for not empowering participants to identify the topic and method of inquiry~\cite{birhanePowerPeopleOpportunities2022,ojewaleAIAccountabilityInfrastructure2024,delgado2023ParticipatoryTurn}. This constrains participant agency and limits the epistemological scope of audits~\cite{lam2022end}. In this section, we describe our long and iterative process of collaborating with a community partner to identify a topic and method of conducting algorithmic audits to illustrate how researchers can scaffold the co-design of algorithmic audits.
In this paper, we draw on 18 months of interaction with our partner organization, a labor union based in Colorado, that culminated in the design and deployment of \Sys{}, which is currently being actively used by rideshare labor organizers. We began the project in June 2023 after an author contacted a partner organization with whom they had a longstanding relationship. We had recently won a grant to build infrastructures that support community-led algorithmic audits in the platform economy and were looking for a partner organization to collaborate with. Our partner was preparing for the 2024 legislative session where they hoped to introduce two bills regulating opacity and dark patterns in rideshare and delivery work. Although they had attempted to pass a bill that regulated transparency in rideshare work during the previous year's legislative session, their effort failed, and they were looking for ways to improve their advocacy in the 2024 session. 

Some organizers thought that the 2023 bill failed to pass because they were under-prepared for bill negotiations with platforms. These organizers believed that legislators viewed platform lobbyists as more authoritative than organizers because they were able to cite internal data to back up their claims. When we approached our partner organization with a proposal for aggregating data about rideshare working conditions, these organizers were excited. They felt that the project could help address their informational disadvantage in bill negotiations with platforms like Uber and Lyft.

Between June to September of 2023, we virtually met with our partner organization several times to build rapport and brainstorm ideas for how an auditing tool could support their goals for inquiry. Organizers and members from the union were broadly interested in questions of fairness in rideshare and delivery work. During these meetings, members from our partner organization shared anecdotes about misleading bonus and quest offers and theories about how platforms intentionally prevent drivers from fulfilling the requirements they'd need to qualify for incentives like bonuses. However, systemically collecting data about offers was prohibitively labor intensive for workers---they would have to screenshot offers as they appeared before the offer expired (usually 15-30 seconds)---and potentially dangerous, as workers often got offers while driving.

\subsection{Co-designing Take Rate as a Topic of Inquiry}
Participatory design traditions call for empowering community members to participate in the overall design of a research study~\cite{epstein2008rise}. We describe how community members and the research team investigated several possible topics of inquiry that were of interest to the community before mutually agreeing on one that operationalized an overall area of interest (e.g., fairness) in a way that was tractable to measure (e.g., take rate).
In late July 2023, an organizer from our partner organization suggested we investigate ``take rate,'' the share of a customer fare that the platform extracted from each ride, as a way of assessing fairness in rideshare work. The organizer had heard from several members who reported that passengers told them they were being charged amounts more than what the drivers were receiving from the platform. During a subsequent meeting with members and organizers from our partner organization, the research team proposed that we follow up on the union members' interest in ``take rates'' and center the inquiry on this topic. Members from our partner organization were enthusiastic. Several recounted instances where an angry passenger got into their car and complained about paying an exorbitant price, which prompted drivers to show the passengers how much they were getting paid for the ride to de-escalate the situation.

Through iterative conversations with organizers and members from our partner organization, we co-designed the topic of inquiry, ``take rate''. This inquiry was more tractable to investigate than offer data and aligned with partner organization interests. The remainder of the summer was spent co-designing the method for inquiry, or how the investigation into take rate would be conducted. At a high level, calculating the take rate requires two primary data points: the amount a customer paid for a transaction and the amount a worker was paid for the same transaction. The research team initially proposed that we could run a small study with our partner organization where drivers would place a QR code flyer in their car that passengers would scan and enter in the amount they paid for a ride. We would record that information and the time of the passenger's submission. Drivers would then enter the amount they were offered for the ride on an interface on their phone. The research team would then link rider responses with driver data by matching a unique ID associated with the QR code in a driver's car and the time at which a submission was made. After proposing this idea to drivers, it immediately became clear that there were problems with this approach.

\subsection{Iterating on a Technical Approach}
The method identified by the research team fulfilled the technical requirements of the project: it would produce data about both the rider price and driver pay. However, through discussion with our partner organization, we quickly realized that just because a data collection method is technically feasible does not mean that it is appropriate.
Union members highlighted that collecting rider data posed significant risks for drivers, including potential deactivation from customer complaints about discussing pay or violating the terms of service (ToS) rules against third-party advertising. They feared such complaints could lead to serious financial consequences as drivers could be deactivated (e.g., fired) as a result of customer complaints. As a result, the research team went back to the drawing board to brainstorm how this data could be collected from riders while minimizing risk to drivers. One of the research team members discovered \textit{Argyle}, a third-party system that branded itself as ADP, a payroll management software, but for non-traditional work.\footnote{Like ADP, Argyle provides access to payroll data but for non-traditional (e.g., non-W2) employees. }A team member discovered the Argyle API provided customer price data and tested it by requesting rides from another team member who was an Uber driver and comparing the data in Argyle with the amount paid and received by the two team members, confirming its accuracy. Following these successful tests, the team proposed using Argyle to investigate take rates.

In August of 2023, the research team proposed using Argyle to collect customer and driver payment data in a way that minimized risk to drivers (e.g., they would not have to facilitate conversations with passengers or hang signs in their cars to know how much a passenger paid). Organizers and members from our partner organization approved this method for conducting the inquiry and voiced satisfaction that their concerns about driver safety in participating in the research study were attended to. Thus, starting in September 2023, the interface design for \Sys{} began. Throughout the interface design process, our partner organization provided regular input into the content and language presented on the \Sys{} interface.

Throughout the remainder of 2023, our partner organization invited us to coalition meetings with other labor unions representing rideshare workers. Through these meetings, we formed relationships with three other labor unions interested in using the FairFare tool to collect wage data with their members. These partner organizations were broadly interested in having access to \textbf{any} type of independent, aggregate quantitative wage data but did not have a specific legislative or organizing goal they wanted to support with data. We supported these organizations' efforts and developed recruitment and onboarding materials to help them sign up their members. In this paper, we focus on our field deployment and evaluation with our original partner organization, as they were our primary co-design partner who played the most active role in co-designing \Sys{}.

\subsection{Research team positionality}
The process of engaging and co-designing with our partner organization was shaped by our positionality as researchers. While we are all HCI scholars, one of us is also a part-time union organizer with our partner organization. Here we describe our positionality in more detail to reflect on how it our study.

Our interdisciplinary team includes experts in HCI, AI, \& Machine Learning, Technology Ethics, and Policy. As academics advocating for platform transparency, we have engaged with gig workers, policymakers, and union organizers. Some of us are involved in designing decentralized platforms to mitigate AI's negative labor impacts, one author is a part-time organizer with our partner organization, and one author works as a part-time Uber driver.
Our positionality influenced data interpretation and participants' engagement. Despite practicing reflexivity and adhering to ethical research practices, we acknowledge potential biases, including sympathy towards organizers and workers as a result of non-transparent AI and algorithmic rideshare platform decisions. These biases likely impacted our data collection and analysis processes. In particular, we likely had response biases~\cite{paulhus1991measurement} in our interviews with organizers when we asked them to reflect on how our system helped them achieve their goals. Organizers were largely uncritical of \Sys{} in their interviews, which could reflect their reluctance to criticize the research team with whom they planned to continue partnering to support their data collection needs. We further unpack organizers' overwhelmingly positive views about \Sys{} and quantitative data more broadly in our discussion.

The process of designing \Sys{} was collaborative and reflexive---we relied on our partner organizations to direct the topic and method of inquiry. As researchers, the initial method we proposed seemed sensible to us. However, our partner organizations' concerns pushed us to reflect on the potential harms of participating in research studies we were not previously aware of. Like with any social computing system, \Sys{} had to fit within our user's everyday routines and expectations to be successfully deployed. In the following sections we describe the technical architecture of \Sys{} (\ref{sec:fairfare-tool}), quantitative data collected during our field deployment with our partner organization (\ref{sec:field-deployment}), and reflections on the process of conducting community-based audits alongside labor unions (\ref{sec:discussion}).

\section{The \Sys{}  System}

\Sys{} is an open source web-based application we designed to minimize friction in driver data sharing while collecting crucial information for researchers and organizers studying rideshare platforms. We developed \Sys{} for the web to maximize device compatibility and allow drivers to sign up and share data without installing additional software. Designing for the web also facilitates deploying custom surveys directly to drivers, which makes the tool more flexible and expands what kinds of data drivers can share.

\subsection{Our Strategic Partner Organization and their Needs}
During the summer of 2023, our research team collaborated with our partner organization to design \Sys{}. Leveraging Argyle, \Sys{} accesses driver wage and rider price data to investigate platform take rates—the percentage of the total customer charge retained by the rideshare platform. The system aligned with an ongoing policy effort in Colorado to address pay transparency and fairness in rideshare work. Through conversations with our partner organization's organizers, we identified the following key design needs: %

\begin{enumerate}
    \item \textbf{Collect data on driver earnings \textit{and} rider cost, to estimate platform take rate.} To calculate take rate, both the consumer cost of a trip and the net driver pay should be collected.
    \item     \textbf{Make driver data donation as seamless as possible.} Organizers mentioned a prior data collection process that asked workers to manually submit screenshots to organizers. If possible, making the data sharing process more automatic would help organizers in their recruitment.
    \item \textbf{Provide access to aggregate data}. Organizers stressed the importance of aggregate data about fare take and pay to better understand the state of driver earnings, test claims made by drivers, and present a defensible perspective to lawmakers. They saw individual drivers benefiting mainly from \textit{``collectivizing''} data for policy advocacy rather than improving individual decision-making. %
    \item \textbf{Enable production of public reports and visuals}. Organizers emphasized that translating any raw or aggregate data into visual summaries of any collected data in the form of graphs and charts were one of the most useful tools for communicating arguments to policy-makers and the public.  

\end{enumerate}

\subsection{Data Protection Strategy and Third-Party Dependencies with Argyle}

Any system that collects data about workers' activities carries inherent risks. Trip data, which is the main data \Sys{} collects, necessarily includes information on work locations and times. Patterns in this data could potentially be used to identify individual drivers, exposing them to retaliation risk by the platform.

We designed \Sys{} to minimize this risk while still enabling meaningful data aggregation for advocacy. We store personal identifying information, such as names, contact details, and account IDs in an isolated table with restricted access. Data on trip details and earnings are stored separately, with row-level security that ensures drivers can only access their own records. Broader data access is only available through an administrator key that is never exposed client-side and not shared outside the research team. Drivers maintain control over their data and can request complete deletion at any time through a contact form in the web app. At the time of deployment this process was handled manually, but it has since been replaced by an automated system that ensures immediate data deletion and un-enrollment from the study.

We rely on a third-party service, Argyle\footnote{https://argyle.com/}, to reliably access platform data. Argyle significantly lowers barriers to participation by making data sharing a simple process for drivers. Argyle's terms of service also help ensure worker privacy in two key ways. First, Argyle's ToS explicitly limits the authorized use of worker data by only accessing and transferring worker's data to specific partners that workers have consented to (in this case, \Sys{}). Second, their terms clearly states that Argyle has ``no relationship" with sources of data like Uber's driver platform. This means that Argyle has no contractual obligations to share information, such as a driver's affiliation with \Sys{}, with data sources like rideshare platforms. \cite{argyle_systems_inc_consumer_2023}. 

However, this makes \Sys{} dependent on their privacy practices and requires an additional layer of trust from drivers—they must trust Argyle as well as the \Sys{} tool. There are two main risks that using Argyle introduces. First, the risk that Argyle may re-sell access to participant data, and second, the risk that the TNC may learn which workers are participating in the study.
TNCs could still potentially discover which workers contributed data through subpoenas or court orders. This risk is reflective of the well-documented difficulties of conducting independent academic research into black-boxed, platform owned systems~\cite{vertesi2023}. In our discussion, we expand on the implications of platform monopolies over data access. Furthermore, we plan to develop our own data collection methods in future work to eliminate this third-party dependency and further reduce privacy risks.

\subsection{How Drivers Interact with \Sys{}}
\label{sec:driver-interaction}

The driver experience consists of driver onboarding, data consent, and initial data collection (also depicted in Figure \ref{fig:driver-interaction}) following these steps:

\begin{enumerate}
    \item Drivers may optionally select an affiliation from a predefined list of organizations or enter a new one.
    \item Drivers review and give consent for data sharing with researchers.
    \item Drivers link their rideshare account to \Sys{}, authorizing access to historical and ongoing work data.
    \item \Sys{} sends a text message with a link to a survey about fare take rates.
    \item After survey completion, drivers view a summary of their data, including the platform's average, highest, and lowest take rates for their historical rides.
\end{enumerate}

\subsection{\Sys{} Architecture}
\begin{figure}[htb]
    \centering
    \includegraphics[width=1\linewidth]{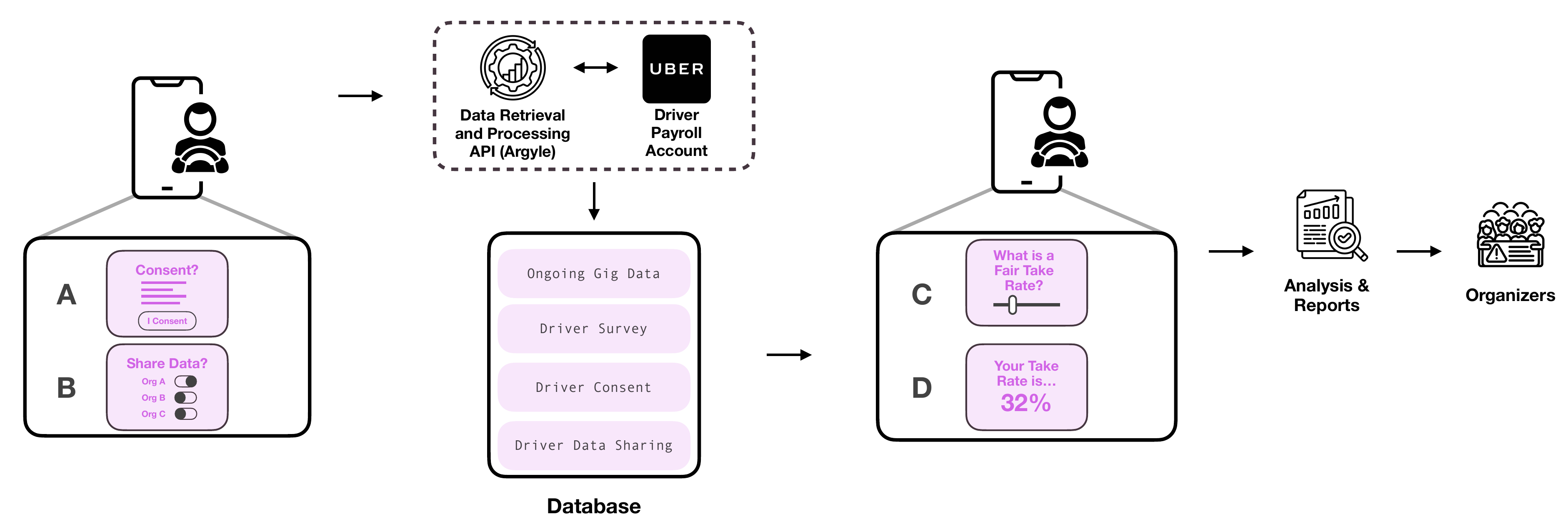}
    \caption{Overview of the \Sys{} flow. Drivers access the \Sys{} through a web browser. They choose whether to provide consent to being a research participant (A), detail their data-sharing preferences (B), and connect their Uber accounts. After connecting an account, the system retrieves historical and future data on workers' gigs. These data are stored in a database alongside additional information like survey answers. Once a minimum amount of data has been retrieved, \Sys{} sends a link to the driver to invite them to complete a short survey (C) and view their average take rate (D). Aggregate driver data are then used to generate reports and analyses for organizers.}
    \label{fig:fairfare-system}
\end{figure}

\label{sec:fairfare-tool}
\Sys{} is a web application built with Svelte\footnote{Svelte is a ``free and open-source component-based front-end software framework'' available at \url{https://svelte.dev/}} and hosted on Netlify with an AWS backend.  Figure \ref{fig:fairfare-system} presents an overview of the system components. The backend processes handle data collection, storage, and processing using the following components:

\subsubsection{Ride Data Collection from Argyle}
We use Argyle's APIs \footnote{Argyle provides programmatic access to various platforms' payroll; learn more at \url{https://argyle.com/}. The data format for data related to gig work can be accessed at: \url{https://argyle.com/docs/api-reference/gigs}}, a third-party service, to retrieve the entire historical driver data from the driver's rideshare account, including paystubs, ratings, vehicles, and payroll documents. Data collection occurs in stages through events and webhooks, with daily  data refreshes. As per the Argyle website, ``when a payroll account [e.g. Uber] is connected..., all available data and documents are retrieved and processed. The information is digitized, standardized, and made accessible..., alongside the original downloadable raw document files''\footnote{\url{https://argyle.com/docs/overview/data-structure/data-sets}}. We restrict to only Uber because at the time of tool development, Lyft data through Argyle did not reveal rider prices, which is required for take rate calculation. Two authors spot-checked Argyle data consistency by requesting several rides where one author was a passenger and another was a driver. Data reported through Argyle's API on these trips matched what both the passenger and driver saw in the Uber app.

\subsubsection{Survey Data Collection from Drivers}
\label{sec:survey-feature}

\Sys{} includes a survey delivery mechanism to help collect additional data from drivers. Surveys are implemented as custom Svelte template files. After a drivers' data is successfully synced, \Sys{} sends them a custom secure link to a survey page. Responses to surveys are linked to driver identifiers and can only be completed once per driver.

This feature is motivated by representatives from our partner organization voicing interest in knowing the difference between driver expectations of take rate and the actual take rate. This survey feature in \Sys{} adds additional flexibility and allows drivers to share data beyond just their historical pay, such as their perspective on platform take rates.

\subsubsection{Data Storage} 
A PostgreSQL\footnote{PostGreSQL is a ``free and open-source relational database'' available at \url{https://www.postgresql.org/}}  database hosted on Supabase\footnote{Supabase is a SaaS backend platform and open source software available at \url{https://supabase.com/}} stores the collected data. Supabase provides an API layer managing authentication and role-based access. \Sys{} includes a light server layer offering an extended API for additional functionality.

\subsubsection{Data Processing} 
\label{sec:data-processing}
A Snakemake-managed pipeline\footnote{Snakemake is a ``workflow management system is a tool to create reproducible and scalable data analyses'' available at \url{https://snakemake.github.io/}} ensures reproducible data cleaning and processing locally through the following steps:
\begin{enumerate}
\item Data export from Supabase. 
\item Merging of user data with metadata, including survey responses and affiliations.
\item Data cleaning and processing:
\begin{enumerate}
\item Calculation of take rate using the formula\footnote{Platform fees is the ``fees charged to the gig employee by the gig platform.'' as per the Argyle documentation} \url{https://docs.argyle.com/api-reference/gigs}. We don't include bonus and other misc payments made or received by the gig employee such as refunds (tolls), taxes, partial cash payments, reimbursements:
\begin{equation*}
\text{Take rate} = \frac{\text{Platform Fees}}{\text{Rider Price - Tips}} \times 100
\end{equation*}
\item Removal of rides with negative take rates. This can occur when platform fees are negative, such as when a rider receives a discount, and the platform incurs a loss.
\item Filtering out non-rideshare activities (e.g., food delivery) and removing entries with ``NA'' values and unnecessary columns.
\end{enumerate}
\item Saving processed data as a Pandas\footnote{Pandas is ``a software library written for the Python programming language for data manipulation and analysis,'' described in detail at \url{https://pandas.pydata.org/}} dataframe \footnote{A data frame is ``a two-dimensional, size-mutable, potentially heterogeneous tabular data,'' described in detail at \url{https://pandas.pydata.org/docs/reference/frame.html}}.

\end{enumerate}

\subsection{Reports}

\begin{figure}[htb]
    \centering
    \includegraphics[width=\linewidth]{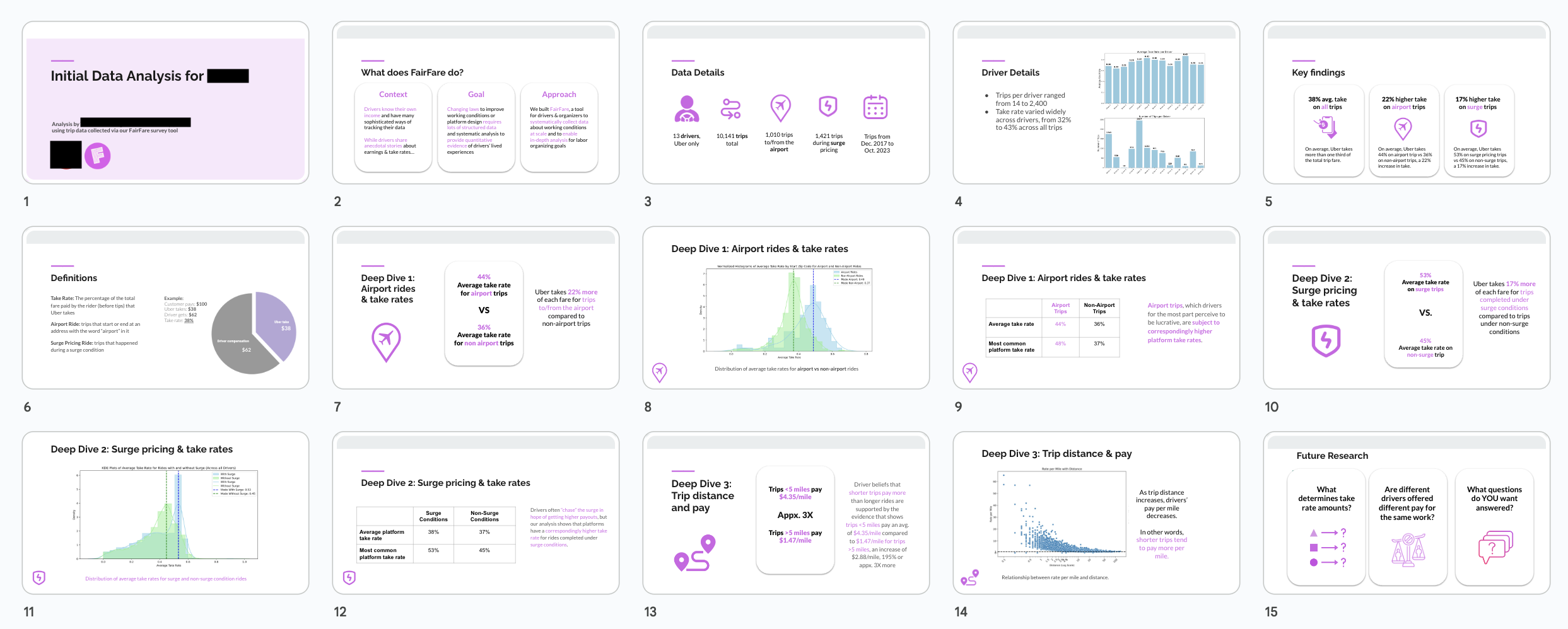}
    \caption{Example report shared with partner organization. This report represents our initial sharing of results based on data collected until October 2023. It contains summary statistics of rides and average take rates. Furthermore it also contains deep dives into how take rate varies for surge vs non-surge rides, airport vs non-airport rides, and how the rate per mile varies by distance.}
    \label{fig:cidu-report}
\end{figure}

Researchers produced reports (see Figure \ref{fig:cidu-report}) for the organizations. We delivered these reports to our partner organization as Google Slide decks. We illustrated each analysis with data visuals and an explainer slide highlighting key takeaways and implications so that even those without a technical background could easily understand it. We chose the plots and figures based on a combination of requests by organizers and our insights to help support advocacy campaigns surrounding the bill. The report depicted here represents an initial sharing of results in October 2023. We present a larger visual of the slide deck in the Appendix.

\section{Field Deployment with Partner Organization}
\label{sec:field-deployment}

Our partner organization circulated recruitment flyers with their members through their email list, WhatsApp group chat, and in-person at the local airport. %
They recruited 45 drivers from whom we gathered data from 76,625 rides from January 2017 to November 2024. On average, rides covered 11.38 miles and lasted 19.36 minutes. The average customer charge was \$24.71, with drivers earning a base pay of \$13.60 plus \$2.82 in tips. The average fee charged was \$7.44, resulting in an overall take rate of 30\%. Table \ref{tab:summary-stats-co} has a summary of the data.

While these aggregate statistics provide valuable insights, it's important to note the evolving nature of Uber's pricing models during our study period. To the best of our knowledge, Uber does not publicly disclose specific timelines or regional details for changes in its pricing models, transitioning from flat commissions to algorithmically variable rates. Uber's own website\footnote{\url{https://www.uber.com/us/en/drive/driver-app/how-surge-works/}} acknowledges that they may test functionality and pricing in ways not described, indicating that different areas could have different pricing models at any given time. Hence our analysis cannot  investigate specific time segmentations and isolate when the commission was a flat fee vs. algorithmically variable. We present aggregate statistics across the entire period of data collection. Based on popular blogs \cite{upfront2022guy}, we believe Uber introduced ``Upfront Pricing'', or a fully AI and algorithmically determined driver wage rather than fixed per mile and per minute rates, to many U.S. markets in early 2022, after a pilot in California in 2019. \footnote{Ubers website states ``Many data points go into calculating an upfront price, including the estimated trip time and distance from origin to destination, as well as demand patterns for that route at that time. It also includes any applicable tolls, taxes, surcharges, and fees (with the exception of wait time fees)''. \url{https://www.uber.com/us/en/ride/how-it-works/upfront-pricing/}}

\begin{table}[htb]
\resizebox{\textwidth}{!}{
\small
\begin{tabular}{@{}l|cr|cccrccc@{}}
\toprule
\multicolumn{1}{c|}{} &
\multicolumn{2}{c|}{\textbf{Total}} &
\multicolumn{7}{c}{\textbf{Average}} \\ 
\midrule
  \textbf{Type} &
  \textbf{Drivers} &
  \textbf{Rides} &
  \textbf{\begin{tabular}[c]{@{}c@{}}Distance\\ (miles)\end{tabular}} &
  \textbf{\begin{tabular}[c]{@{}c@{}}Duration\\ (minutes)\end{tabular}} &
  \textbf{\begin{tabular}[c]{@{}c@{}}Ride Price\\ (\$) \end{tabular}} &
  \textbf{\begin{tabular}[c]{@{}c@{}}Fees\\ (\$)\end{tabular}} &
  \textbf{\begin{tabular}[c]{@{}c@{}}Base Pay\\ (\$)\end{tabular}} &
  \textbf{\begin{tabular}[c]{@{}c@{}}Tips\\ (\$)\end{tabular}} &
  \textbf{\begin{tabular}[c]{@{}c@{}}Take Rate (Average)\\ (\%)\end{tabular}} \\ 
\midrule
All &
45 &
  76,625 &
  11.38 &
  19.36 &
  24.71 &
  7.44 &
  13.60 &
  2.82 &
  30\% \\
  
  Surge &
  37 &
  9,756 &
  10.69 &
  19.28 &
  30.64 &
  10.38 & 
  13.80 & 
  3.36 & 
  31\% \\
  
  Airport &
  34 &
  9,461 &
  26.89 &
  34.40 &
  58.74 &
  17.31 & 
  30.12 & 
  7.54 & 
  29\% \\
\bottomrule
\end{tabular}%
}
\caption{Summary statistics of partner organization's rideshare data Jan 2017 - Aug 2024}
\label{tab:summary-stats-co}
\end{table}

\subsection{Data processing}

To create a final dataset for analysis we first exclude any negative take rates, filter out activities that are non-rideshare, remove activities with any missing data, and exclude `cancelled' rides. This results in a final dataset of 76,625 rides with 45 drivers. We examine several types of rides, and define them as follows. An ``airport ride'' is any ride that starts or ends in the zip code that contains the local major airport. A ``surge'' ride is any ride whose price has been raised due to surge pricing as indicated by a binary flag in our data.\footnote{Surge pricing occurs during spikes in demand and are advertised as more lucrative earning opportunities, see https://www.uber.com/us/en/drive/driver-app/how-surge-works/} To calculate take rate, we use the formula described in our data processing pipeline described in \S \ref{sec:data-processing}, which defines take rate as the ratio between platform fees and the ride price (excluding tips).

\subsection{Take Rate Survey}
As part of the deployment, we developed a short survey in collaboration with organizers to investigate driver perceptions of take rates. Using the survey delivery feature described in \S \ref{sec:survey-feature}, drivers are asked three questions before viewing their personal take rate:

\begin{itemize}
    \item What is your estimate of Uber's average take rate for your fares? (percentage, 0 to 100\%)
    \item What do you think is a fair take rate on your fares? (percentage, 0 to 100\%)
    \item What factors do you think affect your take rate the most? (free text)
\end{itemize}

\subsection{Results}
\begin{figure}[htb]
\centering
\begin{minipage}{0.49\textwidth}
  \centering
  \includegraphics[width=\textwidth]{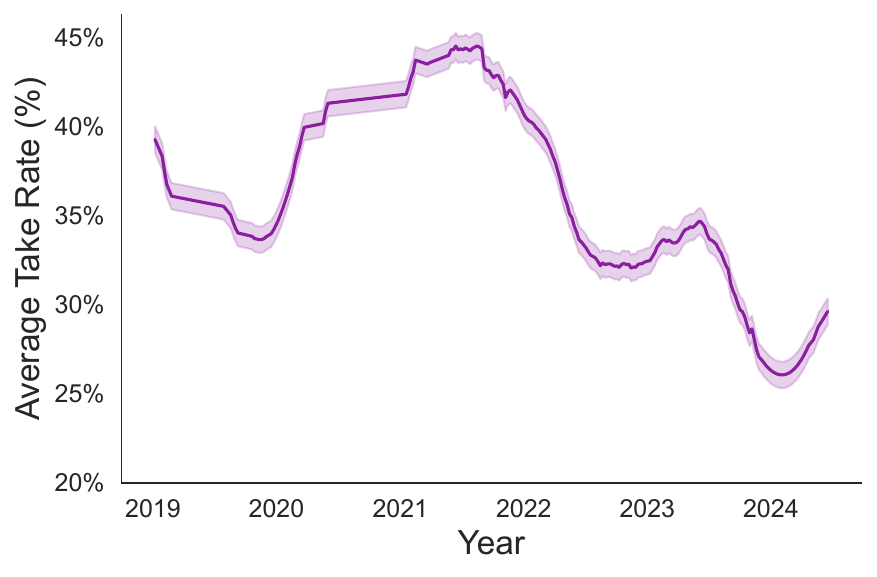}
  \caption{Variation of take rate over time aggregated weekly. The take rate peaks around 2021-2022 and declines steadily through early 2024 before rising again.}
  \label{fig:take-rate-time-co}
\end{minipage}%
\hfill
\begin{minipage}{0.49\textwidth}
  \centering
  \includegraphics[width=\textwidth]{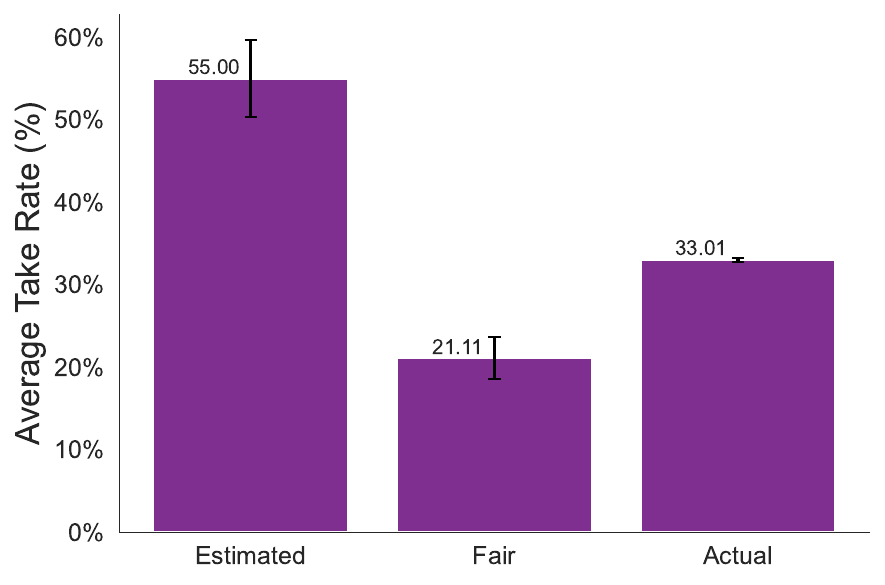}
  \caption{Comparison of actual take rate with driver estimates of current and fair take rates. We find that participating drivers overestimate actual take rates and perceive lower fair rates, with participating drivers' actual rate exceeding their fair rate.
  }
  \label{fig:take-rate-perception-co}
\end{minipage}
\end{figure}

The results reveal significant trends in take rates and driver perceptions across drivers. Figure \ref{fig:take-rate-time-co} shows take rates peaking around 2021-2022 at approximately 44\%, before declining steadily to about 24\% in 2024. Figure \ref{fig:take-rate-perception-co} illustrates a disparity between drivers' perceptions and reality regarding take rates based on their answers to a survey they have to fill out as part of using \Sys{}. Participants overestimated current take rate (55\%) compared to actual rates (33\%). Notably, participants' actual take rate exceeded their perceived fair rate of 21\%.

\begin{figure}[htb]
\centering
\begin{minipage}{0.49\textwidth}
  \centering
  \includegraphics[width=\textwidth]{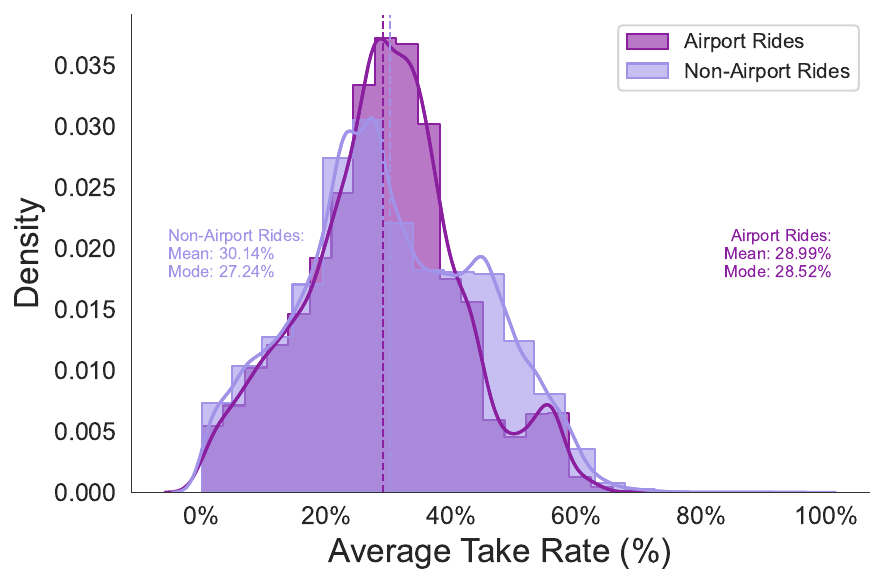}
  \caption{Comparison of take rate between airport and non-airport rides in partner's region. Airport rides, generally preferred by drivers and a common ride type, have a statistically significantly higher take rate *p < 0.05}
  \label{fig:take-rate-airport}
\end{minipage}%
\hfill
\begin{minipage}{0.49\textwidth}
  \centering
  \includegraphics[width=\textwidth]{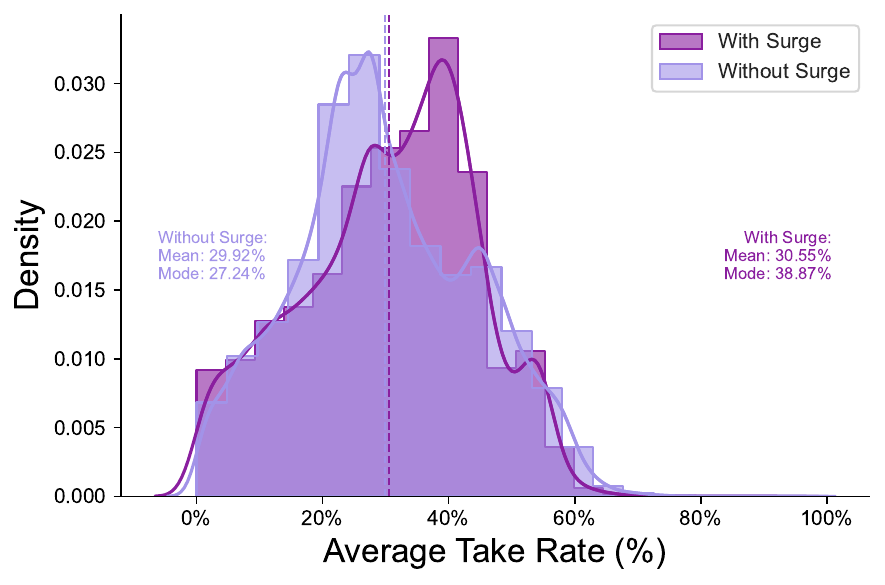}
  \caption{Comparison of take rate between surge and non-surge periods in partner's data. Take rates are higher during surge periods, resulting in a statistically significant difference compared to non-surge periods, reducing driver benefits. *p < 0.05}
  \label{fig:take-rate-surge}
\end{minipage}
\end{figure}

Examining specific scenarios in our partner's dataset, Figure \ref{fig:take-rate-airport} demonstrates that airport rides (9,416 rides or 12\%) most frequently have a statistically significantly (*p<0.05) higher take rate (mode 28.52\%) compared to non-airport rides (mode 27.24\%). Similarly, Figure \ref{fig:take-rate-surge} shows surge periods (9,756 rides or 13\%) most frequently yield statistically significantly (*p < 0.05) higher take rates (mode 38.87\%) than non-surge periods (mode 27.24\%). These findings indicate that the platform captures a larger share of the fare for both airport rides and surge periods compared to non-airport and non-surge trips. In both scenarios, drivers might earn more if the platform maintained take rates closer to those of regular or non-surge rides. This suggests that the potential financial benefits of these desirable ride types (airport trips) or high-demand periods (surges) are not fully passed on to the drivers.

\section{Evaluation with our Partner Organization}

\label{sec:organizer-eval}
To evaluate \Sys{}'s impact on worker organizations' efforts, we conducted a 45-minute semi-structured interview via Zoom with two organizers (An Executive Vice President and Secretary Treasurer) affiliated with our partner organization in summer 2024. 
Our evaluation aimed to address two primary research questions: \textit{(1) How did data access and analysis facilitated by \Sys{} help (or not) in organizing work? How was it different from what they anticipated, if at all?} and \textit{(2) What data or analyses not available in \Sys{} would be most helpful to have in the future?} 

We used thematic analysis to examine the interview transcript. One researcher independently coded the interview and developed a codebook. All researchers then collaboratively reviewed the codes and distilled them into themes, which we present below. The interview protocol is available in Appendix \ref{asec:interview-protocol-cidu-eval}. Our university IRB approved this study.

In framing our findings, we assessed whether \Sys{} met the initial key design objectives of our partner organization outlined: collecting data on driver earnings and rider costs to estimate platform take rates, making data donation seamless, providing aggregate data access, and enabling production of public reports and visuals that could support policy advocacy. While \Sys{} successfully addressed many of these initial goals, organizers also identified opportunities for expansion and continued collaboration.

\subsection{How \Sys{} Reports Helped Organizers Achieve Their Advocacy Goals}

\subsubsection{Organizers Highlighted How \Sys{} Data and Reports Informed Draft Legislation}
\label{sec:bill-language}
Our partner organization organizers discussed how they saw \Sys{} data as useful when drafting a proposed state bill regulating rideshare companies.\footnote{The Governor of Colorado signed the Sentate Bill 24-75 bill into an act on June 5, 2024, compelling rideshare platforms to reveal how rider prices are split between the platform and drivers from February 2025} The bill requires transportation network companies to implement fare transparency requirements and fair deactivation procedures (See Section \ref{sec:policy-outcomes} for a more nuanced discussion and our reflections on actual policy impacts). Organizers viewed the \Sys{} data as a strategic resource for legislative advocacy, articulating how the research \textit{``gave us a clearer picture of what to debate as far as the language in the bill and what would resonate with lawmakers.''} The data also enabled organizers to precisely frame their critiques of rideshare company practices, leveraging empirical data reflecting worker's experiences with compensation. As organizers noted, the data helped them explicitly define \textit{``what our gripes were with rideshare companies,''} transforming anecdotal concerns into evidence-based policy arguments. Moreover, specific data like take rate statistics became critical leverage points: \textit{``the research... helped us demand the need for transparency in our bill.'' (P3)}.

\subsubsection{Data and Reports helped create a national narrative}
Organizers' response suggested a connection between the credibility that \Sys{} reports lent to their work and how they thought the project helped construct a ``national narrative'' across advocacy groups\footnote{\Sys{} gained national visibility in the U.S. through an awareness campaign by a ``progressive non-profit news media organization.'' This exposure led to driver sign-ups nationwide, including interest from union organizers in four other states. As of late November 2024, over 500 drivers nationwide have donated data, resulting in a dataset of over 800,000 rides.}. As organizers highlighted, the reports demonstrated \textit{``that there is a real problem and that this is credible research''}, while also noting the tool's expanding reach \textit{``I've heard your name in different circles just randomly because this research is going in many different people's hands, so it's also building a more national narrative.''} This national narrative refers to the growing awareness and discussion of gig worker issues across various advocacy groups and platforms. Beyond the additional driver sign-ups, the authors of this project have also been been invited to present results and analysis using \Sys{} data at several national gig worker coalition roundtables, where our partner organization organizers have heard the project and results mentioned. These presentations have thus helped connect local advocacy efforts to a broader, nationwide conversation about gig worker rights and fair compensation.

\subsection{How \Sys{} Helped Broader Organizer Data Collection Efforts}

\subsubsection{  \Sys{} Provided Aggregate Data to Challenge Platform Narratives}
\Sys{} transformed the labor-intensive data collection process while equipping organizers with critical insights that rideshare platforms deliberately concealed. Instead of manually gathering thousands of screenshots, organizers could now receive technical reports, as they noted: \textit{``Instead of trying to collect 1000 screenshots and store that...she (a member of our research team) had them in graphs and explained specifically about the what the take rate was in this specific area, how many rides that she used...It was just everything we wanted without having to constantly ask passengers or take screenshots to get it.''}
Platforms deliberately maintain opacity despite their technological capabilities for transparency, as organizers emphasized: \textit{``They could be more transparent in articulating [fare] information to drivers, either in their own language or break it down as simple as possible. And they're so smart and dynamic, shouldn't they know how to do that within the app?''} \Sys{}'s independent data became a powerful tool to challenge this controlled narrative, enabling organizers to educate drivers about their earnings. As organizers explained: \textit{``It would give us the opportunity to educate them on what not to do, or why this doesn't make sense...because these are their earnings. This is their livelihood. So yeah, it would give us more leverage on that front.''}

\subsubsection{\Sys{} Helped Achieve Safe Data Donation}
Without the \Sys{} tool, collecting data on fare take and platform fees could be difficult and treacherous for drivers and organizers. On the driver payroll account for Uber on a desktop, drivers get access to weekly summaries including rider prices; but it is a convoluted and manual process many drivers don't know about, not all drivers have access to this, and rider prices per ride are not available. Organizers also explained that to collect the necessary data for take rate calculations, drivers previously had to talk to passengers and ask them how much they paid for their ride, which could risk deactivation: \textit{``drivers were being deactivated...because they're not supposed to be having those kind of conversations with passengers.''} \Sys{} was \textit{``a pathway to getting that information without the driver being at risk of [the platform] saying he harassed the passenger.''} The design of the \Sys{} tool helped eliminate this risk, making collecting data about drivers easier and less dangerous for drivers.

\subsection{Opportunities for Expanding \Sys{}'s Data Offerings} %
\subsubsection{Need for Broader Driver Participation and More Granular Time Data}
Organizers identified two key areas for improving \Sys{}'s datasets. First, they emphasized the need for wider driver participation: \textit{``It would be great to have a bigger data set in our locale. I think there's quite a bit of data you all have already collected, but it would be nice to have even more drivers.''} This broader participation would provide a more representative picture of driver experiences. However, expanding participation faces challenges, as organizers noted: \textit{``I know part of the problem is just getting people to sign up for it, right? Because some people are wary no matter how many times we say to download this.''} This wariness likely stems from concerns about data privacy and potential platform retaliation as rideshare platforms already collect extensive personal information including location data, payment details and behavioral details, highlighting critical barriers to long-term tool adoption. Moreover, these audits rely on marginalized over-researched populations that are wary of participating in research studies. The wariness expressed in this quote is also likely a manifestation of the overall caution with which the gig worker population approches researchers. Separately, organizers requested more time-series comparisons: \textit{``more specific to the time periods. So like pulling data from 2023 to 2024..."} This granularity would allow organizers to track changes in platform practices more accurately over time.

\subsubsection{Organizers Want to Measure the Effects of Policy Interventions}
Organizers are interested in being empowered to do their own research to evaluate the effectiveness of different policy interventions across jurisdictions. They proposed to \textit{``see overall across the US how they're trending after any kind of legislation has been passed at the state or the local level and see if it's better than those states or cities who have not passed any legislation.''} This approach would provide insights into which legislative approaches truly improve driver conditions, guiding future policy decisions with empirical evidence.

\subsubsection{Need for Long-Term Temporal Analysis}
Organizers stressed the importance of long-term data to track changes in driver compensation and working conditions. P3 and P4 requested historical comparisons, stating, \textit{``I wish we had a comparison chart or something and that the data could start in 2018 right before COVID and then up until now.''} Extending this desire for historical context, they emphasized the need for even earlier data points: \textit{``Because that's when they actually paid a little bit more...my husband was a driver in like 15/16/17 and I know there's no way he would have stood for that as far as the amount they were taking.''} This underscores the need for longitudinal data to quantify perceived declines in driver compensation and analyze changes in platform practices, including responses to major events like the COVID-19 pandemic.

\section{Discussion}
\label{sec:discussion}

Our work with \Sys{} addresses the gap in policy-driven system development and community engagement in HCI research \cite{yang2024future} \Sys{} automates data collection and take rate analysis at scale, providing independent insight into pay, a dimension of working conditions, in the rideshare industry . Organizers were able to leverage FairFare to participate in ``data rhetoric''~\cite{khovanskaya2019} with platforms in bill negotiations–using the same ``data driven'' language that rideshare platforms use to dominate policy and public discourse. We demonstrate how techniques from social computing like algorithmic auditing and co-design can inform the development of tools that go ``back to labor''~\cite{greenbaum1996} to produce evidence of digital technologies' impact on the labor process. These data are valuable for leveling the playing field in policy debates between labor organizers and labor platforms. In the subsequent sections we reflect on our efforts in shaping policy, implications for the design of future transparency tool,s and surface limits of community-supported auditing and data transparency.

\subsection{Complexities of Translating Quantitative Data into Policy Outcomes}
\label{sec:policy-outcomes}
\subsubsection{Legislative Outcomes and Transparency Measures}
Colorado bill SB 24-75\footnote{\url{https://leg.colorado.gov/sites/default/files/2024a_075_signed.pdf}} was signed into law on June 5, 2024, marking a significant milestone in advancing transparency for rideshare drivers and consumers. Our collaboration with organizers played a key role in supporting labor organizers' negotiations with rideshare platforms during the bill drafting process. In particular, organizers told us that they felt more informed and prepared in negotiation sessions because they had independent access to quantitative pay data. While the direct influence of our data on specific provisions remains nuanced, organizers believed it validated their anecdotal evidence and contributed to critical elements of the bill. For instance, organizers viewed examples from \Sys{} data, such as rides with very high take rates (e.g., 80\% or higher), as instrumental in providing quantitative evidence that backed up drivers' anecdotes about unfair pay practices.

The bill mandates that, starting February 1, 2025, Transportation Network Companies (TNCs) must provide transparent payment information to both drivers and consumers, as outlined in Sections (11)(b) and (d). Specifically, TNCs are required to disclose the total amount paid by consumers and received by drivers before tips on a single screen of their digital platforms after consumer dropoff. This provision likely reflects the collaborative efforts of multiple stakeholders, including our research team. Although we cannot trace specific phrases in the bill to our data, organizers felt that \Sys{} data helped substantiate their arguments during negotiations and influenced ``bill language'' (See Section \ref{sec:bill-language}).

\subsubsection{Stakeholder Dynamics in Policy Negotiations}
The legislative process underscores several important points about translating data into policy. First, while legislators adopted transparency measures regarding wages and deactivations, this session did not prioritize discussions on minimum pay standards. However, these transparency requirements lay the groundwork for future conversations about establishing such standards. Second, the absence of specific thresholds for take rates in the bill highlights the complexities of multi-stakeholder negotiations, including platform input on bill text. The final bill text and provisions are products of  multistakeholder negotiations that reflect the desires of both labor organizers and rideshare platforms. Just because labor organizers had access to quantitative data describing pay in rideshare work within Colorado does not mean that they were able to exert dominance in bill text negotiations and insist upon setting specific thresholds. Policy-making is a collaborative process, often involving opinions and input from opposing interest groups.

Finally, although the term ``take rate'' does not appear in the final legislation, the essence of ``take rate'' is preserved. In particular, the bill mandates that platforms report both consumer price and the amount of the consumer price that a driver receives for a ride. These are the two primary data points needed to calculate the take rate. While this specific presentation of data puts the onus on drivers to calculate the percent platforms keep as their ``take rate'', they can now do these calculations easily. Previously, this was challenging --- some drivers could access weekly wage breakdowns but only on the web portal, or they could directly ask passengers, risking the consequences of potential complaints and platform deactivation. 
\subsubsection{Implications for Future Advocacy and Transparency}
Overall, while this legislation leaves room for future policy developments—such as minimum pay standards—it marks an essential step toward greater transparency for all stakeholders involved. Having access to quantitative data is not itself enough to guarantee regulatory wins. However, having access to detailed, aggregate quantitative payment data can help organizers and workers alike establish a baseline measurement of existing financial conditions in the rideshare industry which can inform their advocacy efforts. Designing tools to probe and measure working conditions, like pay, is essential to developing a comprehensive set of demands for change in working conditions~\cite{gallagherDigitalWorkerInquiry2023}

The finalized text in a legislative bill is the product of multi-stakeholder negotiations and reflects the desires and power dynamics of those involved. Our partner organization had many goals for the legislative session, one of which was improving rideshare drivers' everyday working conditions. In this, our partner organization succeeded: drivers will be shown additional information about offers, given more time to process offers before choosing to accept or reject them, be shown consumer prices, and benefit from a formalized and more transparent deactivation process.
\subsection{Design Implications For Transparency Tools}
While \Sys{} helped organizations gather systemic evidence of platform practices, our deployment with our partner organization surfaced important tensions that future tools should carefully consider.
One core tension that \Sys{} attempts to solve is the need to gather data about platform practices and working conditions from an already over-researched working population that can often be wary of institutional engagement.
Data collection architecture needs to balance ease of use with worker trust. Argyle, the third-party data sharing API we used in \Sys, significantly lowered barriers to data sharing. However, it also introduced new trust requirements: workers were required to log into their account through yet another third-party service. Future tools should investigate how to give workers more direct control over data sharing while maintaining similar levels of usability.
Working with trusted labor groups was critical to establish legitimacy and facilitate worker adoption. Deploying \Sys{} with our partner organization helped contextualize the data collection within broader organizing efforts, which we believe helped increase worker trust. However, this also meant that \Sys{} was heavily dependent on this organizational relationship, their resources, and the level of trust workers placed in them. It's also important to consider the broader implications when such systems target specific vendors. For instance, if data is available for one platform (e.g., Uber) but not its competitors (e.g., Lyft), it could lead to asymmetric market dynamics, potentially prompting the targeted platform to restrict data access or advocate for legislation mandating disclosure from all competitors.

\subsection{Limitations and Lessons Learned}
Although \Sys{} successfully bridged worker experiences and policy advocacy, it revealed limitations in using crowdsourced data for legislative change. Increased access to quantitative data does not, in itself, lead to impact. Aggregate data often proved less effective than extreme examples in advocacy. Our work was also limited by reliance on publicly accessible data, with crucial data points still held exclusively by platforms.

\subsubsection{Limits of Community Auditing in Legislative Advocacy}
Data plays only a small role in the overall legislative advocacy process. Throughout the interview process, all of our participants emphasized the importance of data for their movements as they believed it could help them disrupt the monopoly rideshare platforms hold over the public narrative about working conditions in the gig economy. Independent access to platform data was crucial to helping drivers tell their own stories about working conditions. Quantification serves as a legitimizing tool~\cite{scott2020seeing} that helps communities encapsulate their experiences and tell \textit{their} truth~\cite{harris2024}. Access to quantitative data through \Sys{} helped organizers develop independent mental models about working conditions in rideshare which helped them feel more informed during the bill writing process. However, they did not indicate that they used the data elsewhere in their legislative advocacy to achieve their goal of disrupting the public narrative about rideshare working conditions.

Legislative advocacy often requires highlighting extreme examples to demonstrate marked injustice. Data from \Sys{} flattened those extreme examples in averages and were therefore likely unattractive for public campaign use. Data production is a value-laden process~\cite{d2023data,d2024counting}. Organizers wanted data that \textbf{supported} their arguments. For organizers, the purpose of data was to provide another pathway for legitimizing worker voice by tapping into the rhetorical and political power that quantification holds~\cite{scott2020seeing}. When quantitative data did not support their arguments (e.g., the dissonance between expected and actual take rates), organizers sought out ways to produce data that aligned with their members' anecdotal data. All data production efforts reflect the politics of those who are producing them. Understanding when and how organizers choose to use data to selectively support their arguments is an important area of future study. 

\subsubsection{Limits of Data Transparency}
To hold rideshare platforms accountable and enable public oversight of their opaque AI and algorithmic decisions, comprehensive data on algorithmic inputs and outputs is crucial. While this data remains largely inaccessible, \Sys{} takes a step towards enhancing rideshare transparency by focusing on platform take rates. Our approach partners with organizers, crowdsources data from drivers, and ultimately aims to inform policy to improve worker well-being. However, drivers require transparency on various other automated decisions as well, such as promotion qualifications, ride assignments, variations of driver wages and rider prices by demographics~\cite{rao2024transparency}, most of which remain exclusively with the platforms. This data gap limits the scope of research questions that academia and civil society can address.

Public data disclosures, mandated by some cities like Chicago\footnote{\url{https://data.cityofchicago.org/Transportation/Transportation-Network-Providers-Trips-2018-2022-/m6dm-c72p/about_data}} and New York\footnote{\url{https://www.nyc.gov/site/tlc/about/tlc-trip-record-data.page}} in the U.S., represent a positive step but remain insufficient. These disclosures lack critical information such as drivers' part-time/full-time status, driver and rider demographics, and prevent linking qualitative information to individual-level data. To effectively monitor platform behavior, we require more comprehensive public disclosures. This can be made available through recently proposed, periodically publishable, public rideshare transparency reports; and also in an accessible format for researchers, such as through the availability of APIs which may be queried through computer programs~\cite{rao2024transparency, rao2024memo}.

However, relying solely on platforms for data disclosure is not a panacea for accountability. Platforms lack incentives for comprehensive public data disclosures, potentially leading to ``transparency washing'' that obfuscates fundamental questions about power concentration \cite{zalnieriute2021transparency}. Moreover, data disclosure efforts without commensurate analysis, monitoring, and enforcement infrastructures are largely toothless. We echo critiques of the open-source movement to argue that mandating data disclosures from labor platforms is unlikely to precipitate meaningful changes in the structure of this precarious work~\cite{widder2023open,widder2024open}. Data, and access to data, necessitate infrastructures to \textbf{do something meaningful} with it. In the case of \Sys{}, we attempt to scaffold the infrastructure to do something with pay data by creating accessible reports that connect data points to contexts that are meaningful for rideshare drivers and their advocates (e.g., how take rates coincide with popular types of ride offers like airport and surge rides). Our collaborative approach with \Sys{} goes beyond mere transparency by engaging workers and organizers, facilitating data donation, and fostering community engagement. This method aligns with critiques of transparency alone, emphasizing the need to understand platforms through their relational nature with human and non-human elements, leading to sustained accountability and effective governance solutions \cite{ananny2018seeing}.

\subsection{Future Research Opportunities}

The success of \Sys{} in influencing policy and addressing real-world challenges opens up exciting new avenues for HCI research. Future work can build on this foundation in two key directions: expanding \Sys{}'s capabilities and further integrating policy considerations into HCI research methodologies.

\subsubsection{Expanding \Sys{}'s Capabilities}

\Sys{} has already contributed to the passage of legislation benefiting rideshare drivers. Building on this success, the tool could expand its impact in several key areas. These include measuring the effects of newly enacted state-level legislation across the US, supporting legal theories with empirical evidence, and automating existing manual workflows through further tool development and data utilization.

To illustrate these potential expansions with specific examples: \Sys{}'s collected data could quantify real policy impacts by enabling analysis of driver earnings before and after rideshare minimum wage law implementations across states. The data could also provide empirical evidence supporting legal theories about systemic pay disparities. For instance, by combining existing ride location and earnings data with newly collected demographic information, researchers could use \Sys{} data to investigate potential algorithmic wage discrimination. This concept, proposed by legal scholar Veena Dubal, suggests platforms may compensate workers unequally for similar work without adequate justification \cite{dubalAlgorithmicWageDiscrimination2023}. Furthermore, \Sys{}'s data could support the development of new tools to automate hourly wage calculations for specific time periods. These tools could streamline unemployment compensation claims during organizer-run deactivation clinics, helping workers regain platform access more efficiently, and also help drivers plan their schedules in specific locations.

\section{Conclusion}
This work presents \Sys, a tool addressing rideshare worker organizations' data needs to improve algorithmic accountability. Developed collaboratively with our partner organization, \Sys's crowdsourced data on platform take rates directly influenced language in a recently passed transparency state law. This success demonstrates its efficacy in strengthening data-driven policy arguments and addressing calls in the HCI community for tools supporting modern labor organizing. Beyond local impact, \Sys{} has contributed to a broader national advocacy narrative. 
Our findings underscore the potential of collaborative socio-technical infrastructure in addressing algorithmic accountability gaps in the gig economy. Looking ahead, future work could expand \Sys{} to other sectors, enhance its analytical capabilities, and assess its long-term impact on policy and worker well-being.
\label{sec:conclusion}

\section*{Acknowledgements}
The project is partially funded through the Mozilla Technology Fund (2023). We thank Serhiy Ozhibko for helping with design of artifacts.

\bibliographystyle{ACM-Reference-Format}
\bibliography{references}

\appendix

\section{Dashboard}

\begin{figure}[htb]
\centering
  \includegraphics[width=0.6\linewidth]{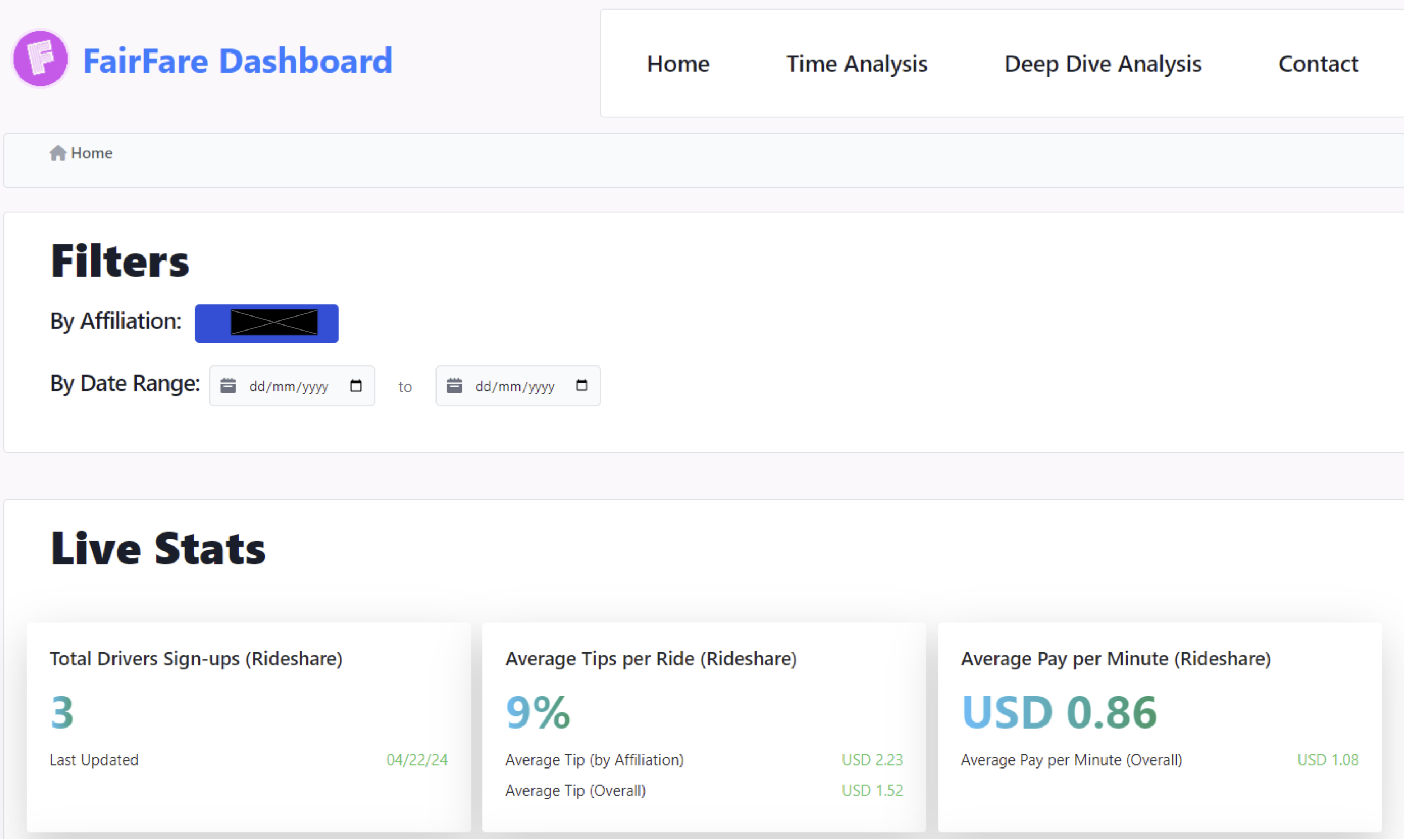}
  \caption{FairFare Dashboard enables researchers to view real-time metrics filtered according to driver affiliations and date ranges}
  \label{fig:dashboard}
\end{figure}%

At the time of initial report sharing, \Sys{} had a dashboard where researchers can access and analyze the collected data (see Figure \ref{fig:dashboard}). For the \Sys{} version used in the Field Deployment described in \S \Ref{sec:field-deployment}, organizers did not have direct access to the dashboard because it was still under active development and used to manage the deployment, not to present results to organizers. Therefore, researchers used the dashboard to get real-time aggregate data and interactive visualizations, including charts and plots like driver income breakdown by categories (e.g. base pay, tips, bonus), and driver perceptions of platform fees. These tools allow for quick insights into the data. Customizable filters enable researchers to refine data based on organizer affiliations and specific date ranges, facilitating targeted analysis. 

For more in-depth analysis, researchers use Jupyter Notebooks\footnote{See \url{https://jupyter.org/}} to work with the cleaned data. The Snakemake-managed pipeline ensures that any changes in the input data or processing scripts trigger updates to the output files, maintaining data consistency and analysis reproducibility.

\section{Interview Protocols}

\noindent \textbf{Introduction}

\noindent Hi everyone, and thank you for coming. Today we’ll be discussing your experiences about \Sys. My name is <> and I am a <>. <> will be our notetakers and observers for this session. You should have received a digital copy of the consent form for this research, which describes the purpose of the study, what we’ll be asking you to do, and your rights as a participant. But I just want to quickly go over it to be sure everyone understands what was in that document and has a chance to ask any questions they might have.The consent form contains the goals of this study and the general idea for today. We will audio record this session, and capture a screenshot of our interaction. We may use a commercial transcription service to transcribe the audio recordings. We will ensure that your personal information remains confidential. We would like to reiterate that your participation is voluntary. You are free to stop your participation at any point. You can continue to participate even if you decide not to answer some questions or participate in certain parts of the conversation. Do you have any questions about the consent document or the purpose of the study? [pause for questions] [when all questions are done or if there are none] Great, it's also totally okay if you have questions about this during the study as well, so feel free to ask any time.

\subsection*{Evaluation of \Sys with our partner organization}
\label{asec:interview-protocol-cidu-eval}

\noindent Today's activity will be divided into 2 main parts:
\begin{itemize}
    \item Assessing the Actual Impact of Data on Organizing
    \item Identifying Gaps and Opportunities for Improvement
\end{itemize}

\noindent There is a possibility that I might cut you off politely, but it’s only for the sake of time.To be sure that we get to everything, there might be some conversations that are cut off, but you can write down things on the scratch paper for us to read later. For the next 2 minutes, let’s do introductions. 

\noindent Let’s get started. [finish intro’s, turn on audio recording]

\noindent \textbf{Part 1: Assessing the Actual Impact of Data on Organizing}
\begin{itemize}
    \item What did you get out of the project?
    \item What was the most and least helpful aspect of the project?
    \item What was surprising?
    \item Now that you've had some experience, how has access to and analysis of data actually helped or hindered your organizing work?
\end{itemize}

Probing Questions:

\begin{itemize}
    \item What metrics or criteria, if any, were used to evaluate the success of \Sys?
    \item If the impact differed from your expectations, why do you think that was the case?
\end{itemize}

\noindent \textbf{Part 2: Identifying Gaps and Opportunities for Improvement}
\begin{itemize}
    \item What information that you haven’t seen in the analysis would be most helpful for your efforts? How would this data help your efforts? How do you KNOW that access to this data is integral to your efforts?
    \item What role do you see the platform should play in such advocacy? (e.g. increase transparency through enhanced data transparency, without the need for tools like \Sys?)
\end{itemize}

Probing Questions:
\begin{itemize}
    \item How would having access to these additional data points make a more compelling case for your demands or strengthen your organizing efforts?

\end{itemize}

\section{Comparison of Statistics between Colorado (CO) and non-Colorado (Non-CO) Drivers}
\label{asec:co-vs-nonco}
\begin{table}[htb]
\resizebox{\textwidth}{!}{%
\begin{tabular}{@{}l|cc|ccccccc@{}}
\toprule
\multicolumn{1}{c|}{\multirow{2}{*}{\textbf{Group}}} & \multicolumn{2}{c|}{\textbf{Total}} & \multicolumn{7}{c}{\textbf{Average}}                \\ \cmidrule(l){2-10} 
\multicolumn{1}{c|}{} &
  \textbf{\# of Drivers} &
  \textbf{\# of Rides} &
  \textbf{\begin{tabular}[c]{@{}c@{}}Distance\\ (miles)\end{tabular}} &
  \textbf{\begin{tabular}[c]{@{}c@{}}Duration\\ (minutes)\end{tabular}} &
  \textbf{\begin{tabular}[c]{@{}c@{}}Customer Charge\\ (\$)\end{tabular}} &
  \textbf{\begin{tabular}[c]{@{}c@{}}Fees\\ (\$)\end{tabular}} &
  \textbf{\begin{tabular}[c]{@{}c@{}}Base Pay\\ (\$)\end{tabular}} &
  \textbf{\begin{tabular}[c]{@{}c@{}}Tips\\ (\$)\end{tabular}} &
  \textbf{\begin{tabular}[c]{@{}c@{}}Take Rate\\ (\%)\end{tabular}} \\ \midrule
CO drivers                                     & 45             & 76,625             & 11.38 & 19.36 & 17.99 & 7.45 & 13.60 & 2.82 & 30    \\
Non-CO drivers                                 & 440            & 758,409            & 7.52  & 15.74 & 16.01 & 6.23 & 12.48 & 2.62 & 28.19 \\ \bottomrule
\end{tabular}%
}
\caption{Summary statistics of CO and non-CO rideshare data Jan 2017 - November 2024}
\label{tab:summary-stats}
\end{table}

\begin{figure}[htb]
\centering
\begin{minipage}{0.49\textwidth}
  \centering
  \includegraphics[width=\textwidth]{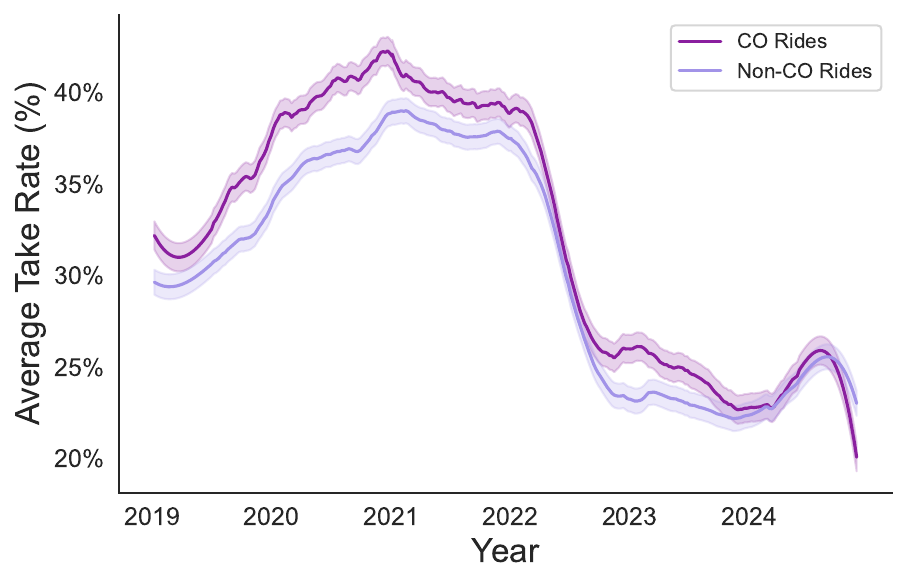}
  \caption{Variation of take rate over time aggregated weekly and its comparison between CO and non-CO drivers. The plot shows similar trends for CO and non-CO ride take rates, peaking around 2021-2022, declining steadily, and converging to similar levels in 2024, with CO rides generally higher.}
  \label{fig:take-rate-time}
\end{minipage}%
\hfill
\begin{minipage}{0.49\textwidth}
  \centering
  \includegraphics[width=\textwidth]{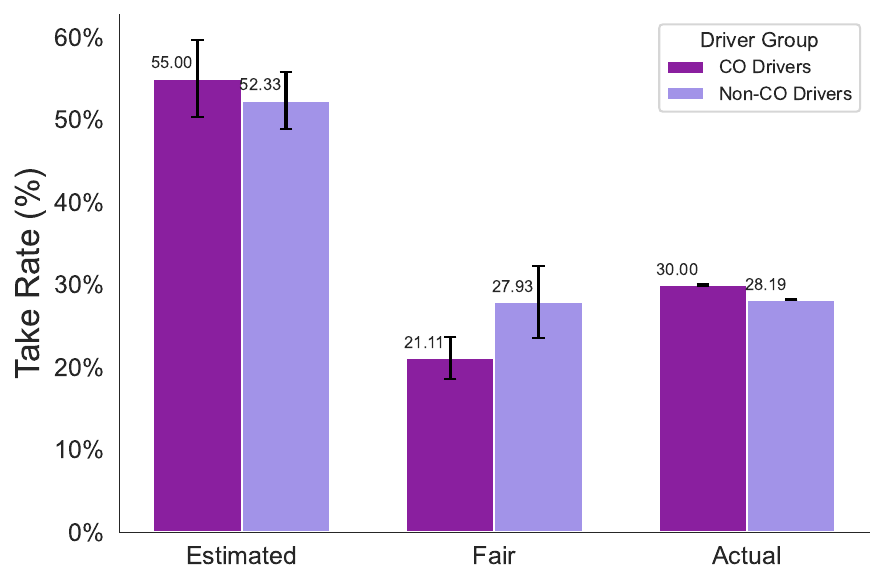}
  \caption{Comparison of actual take rate with driver estimates of current and fair take rates. We find that both CO and non-CO drivers overestimate actual take rates and perceive lower fair rates, with CO drivers' actual rate exceeding their fair rate, while non-CO drivers' actual rate is close to their fair rate.
  }
  \label{fig:take-rate-perception}
\end{minipage}
\end{figure}

\clearpage

\section{Reports}

\begin{figure}[p]
\centering
\rotatebox{90}{%
\includegraphics[width=7in]{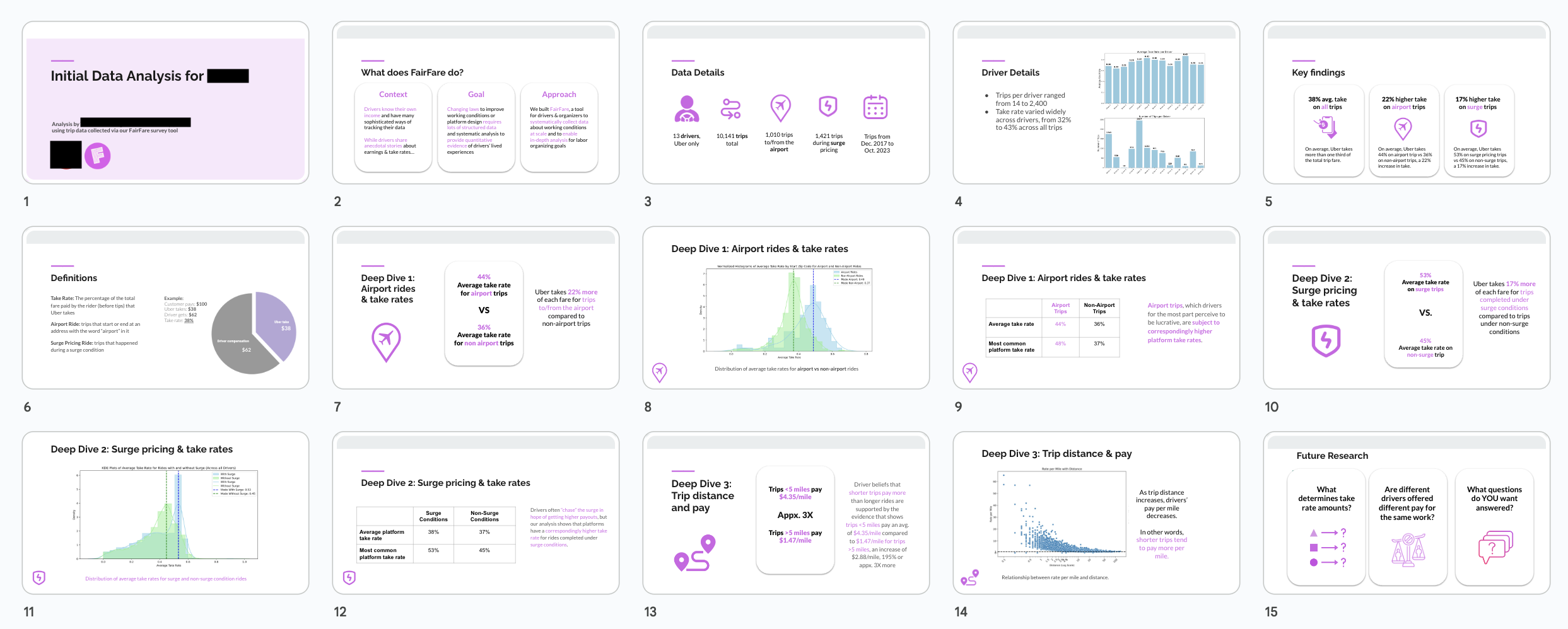}
}
\caption{Example report shared with partner organization. This report represents our initial sharing of results based on data collected until October 2023. It contains summary statistics of rides and average take rates. Furthermore it also contains deep dives into how take rate varies for surge vs non-surge rides, airport vs non-airport rides, and how the rate per mile varies by distance.}
\label{fig:cidu-report-big}
\end{figure}

\end{document}